\documentclass[twoside]{article}
\usepackage{epsfig,fleqn,espcrc2}

\newcommand{\bt}{\beta} 
\newcommand{\kp}{\kappa}

\def\et{{\it et al.}}

\def\prd#1{Phys.\ Rev.\ {\bf D#1}}

\def\npb#1{Nucl.\ Phys.\ {\bf B#1}}

\title{$f_B$ for Various Actions : Approaching the Continuum Limit
with Dynamical Fermions} 

\author{C.\ Bernard\address{Department of Physics, Washington
University, St. Louis, MO 63130, USA}, S.\ Datta\address[IU]{Department of
Physics, Indiana University, Bloomington, IN 47401,
USA}\thanks{presented by S.\  Datta}, C.\ DeTar\address{Physics Department,
University of Utah, Salt Lake City, UT 84112, USA}, Steven\ 
Gottlieb\addressmark[IU], U.M.\  Heller\address{CSIT, Florida State 
University, Tallahassee, FL 32306-4120, USA},
J.\  Hetrick\address{Department of Physics, University
of the Pacific, Stockton, CA 95211, USA}, C.\  McNeile\address{Dept.\  of
Math.\  Sci., University of Liverpool, Liverpool, L69 3BX, UK},
K.\  Orginos\address[AU]{Department of Physics, University of Arizona,
Tucson, AZ 85721, USA}, R.\  Sugar\address{Department of Physics,
University of California, Santa Barbara, CA 93106, USA} and
D.\  Toussaint\addressmark[AU]}

\begin{document}

\begin{abstract}

We present results for pseudoscalar decay constants of heavy-light
mesons using both quenched and $N_f=2$ dynamical fermion configurations.
A variety of fermion actions is investigated:  Wilson, nonperturbative
clover, and fat-link clover. For heavy quarks the Fermilab formalism 
is applied.
In the quenched approximation, results with the nonperturbatively improved
clover action of the Alpha collaboration allow us to study the
systematic error of the continuum extrapolation from the Wilson action.
In addition, we use quenched configurations to explore the effects of 
fattening.  The lessons from the quenched analyses
are then applied to data with dynamical fermions, where both Wilson
and fat-link clover actions have been used.  This allows us to
attempt a continuum extrapolation of the dynamical results.

\end{abstract}

\maketitle

Decay constants of the $B$ and $B_s$ mesons are essential for accurate
determination of parameters of the CKM matrix. We present here 
lattice studies of decay constants of heavy-light mesons.
A MILC collaboration computation of the quenched decay constants 
has appeared previously \cite{prl};
in that work, dynamical simulations were used only to
estimate the quenching error.
Our first continuum-limit results for dynamical lattices with $N_f = 2$ 
staggered quarks were 
presented at Lattice '99 \cite{lat99}. Here the dynamical lattices are
studied in more detail, with both Wilson and clover 
valence fermions. We also update the quenched lattice results, based 
on some new runs and revised analysis. Some of these new runs 
use clover fermions with different levels of fattening (including
no fattening, the standard ``thin'' case), thereby enabling us to
understand better both the fat-link clover dynamical results and the
quenched discretization errors. 

Table \ref{tbl.lattices} gives the lattice parameters. The sets J and
CP are new additions to the quenched lattices previously analyzed in 
Ref. \cite{prl}. Unimproved Wilson valence fermion propagators were
generated for all these sets except J. On J, and a 200-lattice subset
``CP1'' of CP, we used thin link clover fermion propagators. 
For a 100-lattice subset ``CP2'' of CP1, we generated clover
propagators with four different levels of fattening (see below). For
the dynamical lattices, valence Wilson fermions were studied
on all the lattices; fat link clover fermions were also studied 
on a 98-lattice subset ``RF'' of set R, with 10 levels of fattening
and coefficient $c=0.45$.

For our central values, the dynamical quark configurations are 
treated as fixed 
backgrounds and chiral extrapolation is performed in the valence quark
mass only. To estimate the systematic error due to this 
``partial quenching,'' we also extrapolate with $m_{\rm valence}
= m_{\rm dynamical}$, the equality being defined by parameters 
corresponding to the same $m_\pi$. 

\begin{table}[htb]
\vspace{-0.6cm}
\caption{Lattice parameters. The upper group corresponds to quenched
lattices; the lower group, to dynamical lattices with $N_f = 2$
staggered quarks. The set G was generated by HEMCGC.}
\label{tbl.lattices}
\begin{tabular}{clcc}
\hline
name & $\bt (a m_q)$ & size & $ \# $ configs. \\
\hline
A & 5.7 & $8^3\times48$ & 200 \\
B & 5.7 & $16^3\times48$ & 100 \\
E & 5.85 & $12^3\times48$ & 100 \\
C & 6.0 & $16^3\times48$ & 100 \\
CP & 6.0 & $16^3\times48$ & 305 \\
J & 6.15 & $16^3\times48$ & 200 \\
D & 6.3 & $24^3\times80$ & 100 \\
H & 6.52 & $32^3\times100$ & 60 \\
\hline
L & 5.445 (0.025) & $16^3\times48$ & 100 \\
N & 5.5 (0.1) & $24^3\times64$ & 100 \\
O & 5.5 (0.05) & $24^3\times64$ & 100 \\
M & 5.5 (0.025) & $20^3\times64$ & 199 \\
P & 5.5 (0.0125) & $20^3\times64$ & 199 \\
G & 5.6 (0.01) & $16^3\times32$ & 200 \\
R & 5.6 (0.01) & $24^3\times64$ & 200 \\
S & 5.6 (0.02) & $24^3\times64$ & 202 \\
T & 5.6 (0.04) & $24^3\times64$ & 201 \\
U & 5.6 (0.08) & $24^3\times64$ & 202 \\
\hline
\vspace{-1.0cm}
\end{tabular} 
\end{table}  

For clover fermions, both thin and fat,
the full Fermilab formalism \cite{fnl} is applied, 
including use of the EKM norm $\sqrt{1 - 6 \kp}$, the 3-d rotation 
of the fields, and the identification of kinetic mass $m_2$ (rather than 
the pole mass $m_1$) as the physical mass. The shift $m_1\to m_2$ is performed
at tadpole improved tree level. In the thin-link case, we take
the clover coefficient $c_{\rm SW}$ and the light-light
renormalization/improvement constants 
from the 
nonperturbative determinations by the Alpha collaboration \cite{alpha}.
For the heavy-light renormalizations around the B meson mass, we use 
either the one-loop perturbative calculations \cite{ioy} or a version
of tree-level tadpole improvement designed to smoothly
join on to the nonperturbative results as the quark mass gets small.
The former approach is designated ``NP-IOY;'' the latter, ``NP-tad.''
For more details on NP-tad, see Ref.~\cite{claude}. In both cases,
the renormalization of the 3-d rotation (``$d_1$'') term is
performed perturbatively using \cite{ioy}. For the D meson, only NP-tad is 
used since the approximations  in \cite{ioy} are not
applicable for smaller masses.

In the fat-link clover cases, $c_{\rm SW}$ is
set equal to the tadpole improved tree level value $1/u_0^3$. The light-light
renormalization coefficients 
are taken from the perturbative calculations of Ref.~\cite{pert}.
The heavy-lights (for which perturbative calculations do not exist)
are normalized using the {\it static-light} results of \cite{pert}.  
This should be roughly correct for the large values of $aM$ at the B meson.

For Wilson valence quarks, we again use the EKM norm and identify
$m_2$ as the physical mass.  However since the magnetic and kinetic
masses are not equal in this case \cite{fnl}, there is little point in
including the 3-dimensional rotations, and we do not.  The associated
systematic errors are estimated as in \cite{prl}.
We normalize the heavy-light axial current as a function of mass
perturbatively \cite{kuramashi}. For the central values, the
static-light value of the scale, $q^*_{\rm SL} \approx 1.43/a$ \cite{pert} 
is used. (This scale is different from that quoted in Ref.~\cite{hill}
and used by us previously \cite{prl};
for further discussion see \cite{claude}.)

For the light quarks, 3 -- 5 $\kp$ values are used, in the range 0.7 --
2.0 $m_s$. To get to physical light quark masses, we extrapolate
$m_\pi^2$ vs.\ $m_2$ quadratically and 
$f_\pi, m_{Qq}$ and $f_{Qq}$ vs.\ $m_2$ linearly. The systematic error
of the chiral extrapolation is estimated by comparing with
results from quadratic extrapolations for all the
above. For heavy clover quarks, we use 3 -- 5 $\kp$ values 
(giving meson masses in the range 1.8 -- 5 GeV); in the Wilson
case we have $\sim\!10$ heavy quark masses 
as well as static heavy quarks.
Continuum extrapolation is done using both constant 
and linear fits, and the spread is taken as an estimate of the
systematic error. 

\begin{figure}[t]
\epsfxsize=7.0cm
\epsfysize=7.0cm
\epsfbox{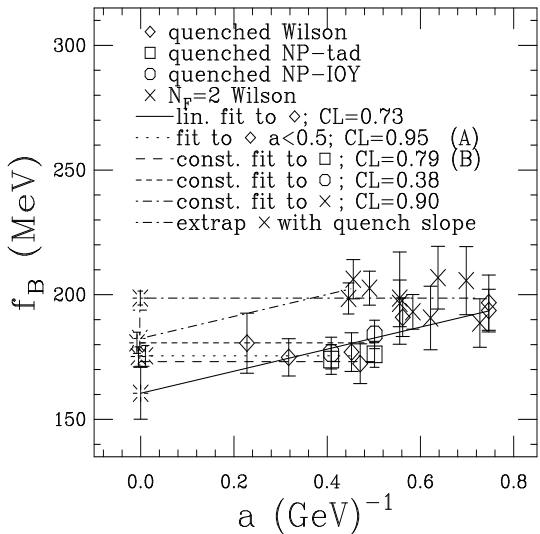}
\vspace{-1.1cm}
\caption{The complete set of $f_B$ data, for both quenched and
dynamical lattices, showing the different continuum
extrapolations. The line for fit (B), which is very close 
to the line for fit (A), has been shifted downward slightly for clarity.} 
\label{fig:fb}
\vspace{-0.9cm}
\end{figure}

Figure \ref{fig:fb} shows the continuum
extrapolations of our full set of data for $f_B$, for both quenched
and dynamical lattices. Both the NP-IOY and NP-tad quenched  calculations,
which should have small discretization errors, are extrapolated with constants.
The results are consistent with both the linear and constant
quenched Wilson extrapolations, although they favor the
constant extrapolation.
All four extrapolations are averaged for our quenched central values;
the spread determines our discretization error.

The dynamical
lattice data clearly favor a constant fit; the best linear fit has
a tiny slope. However, to be conservative, we
also make a linear extrapolation from the data for the smallest $a$,
with the same slope as the linear fit for the quenched set.  
Other systematic errors, including the extrapolation in $1/M_{Qq}$, finite 
size errors, effects of excited states, higher order
perturbative corrections, {\it etc.}, 
are estimated in basically the
same way as in Refs.~\cite{prl,lat99}. However the range of $q^*$ values
considered for the perturbative error has changed along 
with the change in the central value of $q^*$.  We now consider
$1/a \le q^* \le 2q^*_{SL}$, where $q^*_{SL}$ 
is the value calculated in \cite{pert}.

\begin{figure}[thb]
\epsfxsize=7.0cm
\epsfysize=8.0cm
\epsfbox{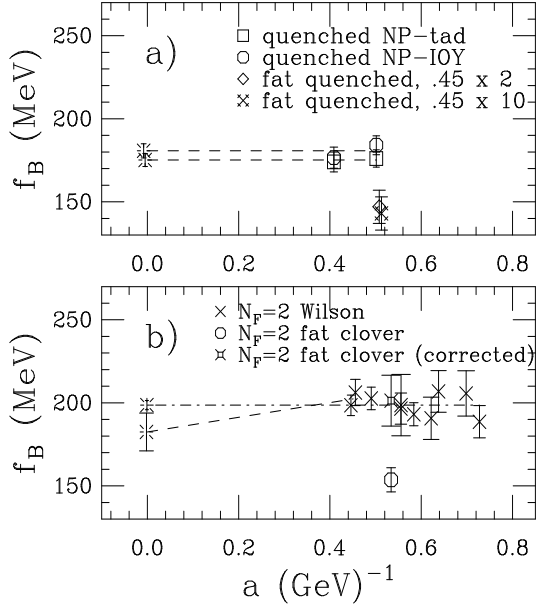}
\vspace{-1.1cm}
\caption{a) $f_B$ for clover fermions on quenched lattices. b) $f_B$ 
on dynamical lattices, for both Wilson and fat-link clover fermions.
Also shown is the ``corrected'' fat-link result.
% by correction factor obtained from a) above.
}
\label{fig:smearing}
\vspace{-0.5cm}
\end{figure}

As reported in Ref.~\cite{lat99}, the fat-link clover results for 
decay constants $f_{D,D_s,B,B_s}$ on dynamical lattices were found 
to be much smaller than the apparent continuum limit results of 
the Wilson quarks, thus raising the possibility that at least one 
of these results had a large uncontrolled systematic error.
To study this issue,
we have computed with clover fermions on quenched
lattices (set CP2) with four different levels of fattening:
links fattened 2, 6 and 10 times with a fattening 
coefficient c=0.45, and links fattened 7 times with c = 0.25.
A comparison with the thin-link clover computations (from sets CP1 and F) 
is shown in Fig.~\ref{fig:smearing}a. 
The fat-link results are considerably suppressed compared to those from 
the thin links. As mentioned above, the latter are consistent with the results
for
continuum-extrapolated quenched
Wilson fermions.
Furthermore, in an unpublished direct investigation of the static $q \bar{q}$
potential, we found that the form of the $q \bar{q}$ potential at 
short distances 
is distorted by fattening. Figure \ref{fig:smearing}a 
also reveals the interesting fact that even the lowest level of
fattening studied by us, {\it viz.}, 2 levels of smearing with a fattening
coefficient of 0.45, already 
substantially suppresses the decay constant. In fact, there is
not much difference in the values of the decay constants between
the four different levels of fattening we studied.

The fat-link studies on quenched lattices, where the approach to the
continuum limit is in much better control, allow us to attribute the
cause of the discrepancy between Wilson and fat clover results on
dynamical lattices to fattening. We note however that the fattening
effects in our calculation include not only the change in the 
short-distance potential, but also the possible error in using
the static-light rather than heavy-light renormalization.
Be that as it may,  the quenched studies clearly show that we
should not average the dynamical fat clover and
Wilson results (as we did in Ref.~\cite{lat99}); instead,
we use the Wilson values only. Interestingly, 
correcting the fat-clover dynamical-lattice result by 
a factor obtained from the quenched lattices 
at comparable lattice spacing
produces agreement with the Wilson fermion 
results (Fig.~\ref{fig:smearing}b), indicating that
the suppression of decay constants due to fattening has 
similar origins for the dynamical and quenched lattices.

Our final results for heavy-light decay constants in the quenched
approximation are (in MeV for the decay constants):
\begin{eqnarray*}
f_B \!\!\!\! & = & \!\!\!\! 173 \pm 6 \pm 16;
\ \ f_{B_s} \! = 199 \pm 5 {}^{+23}_{ -22}; \\
f_D \!\!\!\! & = & \!\!\!\! 200 \pm 6 {}^{+12}_{ -11};
\ \ \ \ f_{D_s} \! = 223 \pm 5 {}^{+19}_{-17}; \\
{f_{B_s} \over f_B} \!\!\!\! & = & \!\!\!\! 1.16 \pm 0.01 \pm 0.02; \\
{f_{D_s} \over f_D} \!\!\!\! & = &\!\!\!\! 1.13 \pm 0.01 \pm 0.02.
\end{eqnarray*}
The errors are statistical and systematic (within the quenched
approximation), respectively. The results differ from those in
Ref.~\cite{prl} due to:

1) The inclusion of new data from sets CP, CP1 and F. 

2) Setting the central value of
the heavy-light scale as $q^*_{SL}$ from \cite{pert},
rather than from \cite{hill}.

3) Other changes in analysis, motivated by the new runs.
These include the error estimate for the chiral extrapolation
(some alternative fits used previously are excluded by the new data)
and the central value of the continuum extrapolation (averaging
our four possible versions 
rather than taking only the linear Wilson fit).

Our results for the dynamical lattices are (in MeV for decay
constants):
\begin{eqnarray*}
f_B \!\!\!\!&=&\!\!\!\! 191 \pm 6 {}^{+24}_{-18} {}^{+11}_{-0};
\ \ f_{B_s} \! = 217 \pm 5 {}^{+33}_{-29} {}^{+9}_{-0}; \\
f_D \!\!\!\!&=&\!\!\!\! 215 \pm 5 {}^{+17}_{-15} {}^{+8}_{-0};
\ \ f_{D_s} \! = 241 \pm 4 {}^{+32}_{-31} {}^{+9}_{-0}; \\
{f_{B_s} \over f_B} \!\!\!\!&=&\!\!\!\! 1.16 \pm 0.01 \pm 0.02 \pm 0.02; \\
{f_{D_s} \over f_D} \!\!\!\!&=&\!\!\!\! 1.14 \pm 0.01
{}^{+0.02}_{-0.03} \pm 0.02.
\end{eqnarray*}
Here the errors are statistical, systematic within $N_f = 2$, and
systematic (due to partial quenching and missing the virtual strange
quark), respectively. The last error is taken to be the largest of
four components : 1) half the difference of the quenched and the $N_f = 2$ 
results (see Ref.~\cite{lat99}), 2) the spread obtained from setting the
scale by $m_\rho$ instead of by $f_\pi$, 3) for quantities involving the
strange quark, the change in fixing $\kp_s$ from $m_\phi$ instead of the
pseudoscalars, 4) the difference from the ``full unquenching'' result 
with $m_{\rm valence} = m_{\rm dynamical}$. For the
individual decay constants, 
errors 1) and 2) are the largest; for the ratios, however, errors 3)
and 4) dominate.


\begin{thebibliography}{9}
\bibitem{prl} C.\ Bernard {\em et al}, Phys.\ Rev.\ Lett.\ {\bf 81} 
(1998) 4812.
\bibitem{lat99} C.\ Bernard {\em et al}, Nucl.\ Phys.\ B (Proc.\ Suppl.)
{\bf 83-84} (2000) 289.
\bibitem{fnl} A.\ El-Khadra {\em et al}, Phys.\ Rev.\ D {\bf 58} (1998) 014506.
\bibitem{alpha} M.\ L\"uscher, \et, \npb{491} (1997) 323;
{\it ibid}, 344. See also R.G. Edwards, U.M. Heller and T.R. Klassen,
Phys. Rev. Lett. {\bf 80} (1998) 3448.
The constant $b_A$
is the perturbative value (S.\ Sint and P.\ Weisz, \npb{502} (1997) 251)
with $q^*=1/a$ (which reproduces  
the nonperturbative value for $b_V$).
\bibitem{ioy} K-I. Ishikawa, T. Onogi and B. Yamada, Nucl. Phys. B 
(Proc. Suppl.) {\bf 83-84} (2000) 301, and private communication.
\bibitem{claude} C. Bernard, these proceedings.
\bibitem{pert} T. DeGrand and C. Bernard, Nucl. Phys. B (Proc.\ Suppl.)
{\bf 83-84} (2000) 845, and in preparation.
\bibitem{kuramashi} Y. Kuramashi, Phys. Rev. D {\bf 58} (1998) 034507.
\bibitem{hill} O.\ Hernandez and B.\ Hill,
\prd{50} (1994) 495.
\end{thebibliography}
\end{document}